# The multiplet structure of the genetic code, from one and *small* number

**Tidjani Négadi**

Département de Physique, Faculté des Sciences,
Université d'Oran, 31100, Oran, Algérie
Email : tnegadi@gmail.com
Website : http://negadi.webs.com

**Abstract**
In this short paper, we show that the multiplet structure of the standard genetic code is derivable from the total number of nucleotides contained in 64 codons, 192, a *small* number. The degeneracy class-number is derived as the number of numbers coprime to the number of Family-Boxes involved for the quartets, the doublets and the singlets. Those for the triplet and the sextets are computed as simple linear combinations of the preceding ones. Some interesting consequences are also presented.
.

## 1. Introduction

The genetic code is a unique system because it shows a clear decimalization, that is arithmetic in base-10. A pioneering work by shCherbak (shCherbak, 1993) has shown that there exist many arithmetical regularities and that the decimal place-value numerical system and also an acting zero are used by the genetic code (shCherbak, 2008). Other people have found similar numerical phenomena (see for example Rakočević, 2009). Some few years ago (Négadi, 2007), we have designed an arithmetic model of the genetic code based on the number 23!, the order of the symmetric group $S_{23}$ of permutation of 23 objects, 20 amino acids and 3 stops. These latter were put in a one-to-one correspondance with the 23 decimal digits of the number 23! and the structure of the five amino acids multiplets as well as the degeneracy of these followed, not to speak about interesting applications (Négadi, 2008, 2009). In this note, we shall pursue this line of thought and give another example where decimalization occurs. We start from the number of nucleotides, 192, composing the 64 codons of the genetic code table and derive the number of amino acids in five degeneracy classes. As the main result in this paper, we show that the computation of the degeneracy class-numbers are derivable as Euler's φ-functions of the number of "family-boxes" involved in the total number of amino acids in each degeneracy class.

## 2. 20 amino acids and 3 stops *from* 64 codons!

The *standard* genetic code is a mapping from 64 coding codons to 20 amino acids and 3 stop codons. On the other hand, each one of the 64 (RNA) codons is a triplet of nucleotides, that is three nitrogenous bases among U, C, A and G. In this way, there are 192 nucleotides in all (64×3). As mentioned in the introduction, we start from the *sole* total number of nucleotides, 192, written in the usual decimal place-value numbering system (base-10). The prime-factorization, using the Fundamental Theorem of Arithmetic writes

$$192=2^6\times 3 \qquad (1)$$

Now, we define the $A_0$-function to be the sum of the usual arithmetic $a_0$-function, the sum of the prime factors (including multiplicity), and the sum of the Prime Indices of these prime factors. (the Prime Index of a prime is an associated natural number counting it; for example PI(2)=1, PI(3)=2, PI(5)=3, and so on.) We have immediately

$$A_0(192)=23 \qquad (2)$$

This result is interesting because it could describe, at the same time, either 20 amino acids and 3 stops, or 23 Amino Acids Signals (see below and the closing remark at the end). In the first case, the 14 digits composing $A_0$ could be sorted as the following multiplet pattern

$$\begin{array}{c} 2+2+2+1+1+1=9 \text{ "doublets"} \\ 2+2+1=5 \text{ "quartets"} \\ 1 \text{ "triplet"} \\ 2 \text{ "singlets"} \\ 3 \text{ "sextets"} \\ 2+1=3 \text{ "stops"} \end{array} \qquad (P)$$

We have therefore the 5 degeneracy classes or 5 multiplets (see the Table below): 5 quartets (G, A, V, T, P), 9 doublets (F, Y, C, H, Q, N, K, D, E), 3 sextets (S, L, R), 1 triplet (I) and 2 singlets (M, W) of amino acids and 3 stops or stop codons (the corresponding codon Table in terms of U, C, A and G could be found for example in Négadi, 2009). To see the link of the above sorting with the distribution of the 20 amino acids in the genetic code 8×8-codons Table, let us look attentively at the following Table where the amino acids are given in the 1-letter well-known code. Note, first, that the 8×8 Table could also be seen as a 4×4 table of 16 "family-box" (FB) indicated by a thick grid. In this way, and importantly for the following, the 5 quartets occupy exactly 5 FBs (as 2+2+1), the 9 doublets in light grey as 2+2+2+1+1+1, see (P), occupy 6 FBs, with sharing, and the 2 singlets 2 FBs, each one of them being shared. Also, the triplet I shares 1 FB with the singlet M and the 3 sextets occupy in all five FBs, one of them shared by a doublet. This "family-box" will prove very fruitfull in the following, concerned with the computation of the degeneracy class-numbers. Before proceeding, let us introduce famous Euler's φ-function of a number n (also called totient function) which gives the number of positive integers not exceeding n and relatively prime to n. (This function, invented by Euler long time ago, plays a prominent role in modern Cryptography.) As a nice and simple result, we have that *the degeneracy class-numbers for the quartets, the doublets and the singlets (16 amino acids) are equal to the φ-functions of the number of FBs invol*ved. Quantitatively, we have

$$5 \text{ Quartets} \rightarrow 5 \text{ FBs} \rightarrow \varphi(5)=4 \qquad (3)$$

$$9 \text{ Doublets} \rightarrow 6 \text{ FBs} \rightarrow \varphi(6)=2=\varphi(\varphi(9)) \qquad (4)$$

$$2 \text{ Singlets} \rightarrow 2 \text{ FBs} \rightarrow \varphi(2)=1 \qquad (5)$$

**Table 1**: Distribution of the 20 amino acids in the genetic code table

| F | F | S | S | L | L | P | P |
|---|---|---|---|---|---|---|---|
| L | L | S | S | L | L | P | P |
| Y | Y | C | C | H | H | R | R |
| s | s | s | W | Q | Q | R | R |
| I | I | T | T | V | V | A | A |
| I | M | T | T | V | V | A | A |
| N | N | S | S | D | D | G | G |
| K | K | R | R | E | E | G | G |

These are exact results: the quartets are made of four codons, the doublets two codons and the singlets one codon. Note that for the quartets and the singlets the number of amino acids and the number of FBs coincide. In the case of the doublets however one has to apply the φ-function to the number of amino acids *two times,* one time to get the correct number of FBs and the second to reach the correct number of codons 2. Finally, it is immediate to get the number of codons for the triplet and the three sextets by considering the ormer as "1 doublet+1 singlet" and the latter as "1 doublet+1 quartet":

$$\text{Triplet: } \varphi(6)+\varphi(2)=2+1=3 \qquad (6)$$

$$\text{Sextets: } \varphi(6)+\varphi(5)=2+4=6 \qquad (7)$$

The Table below summarizes the results

|          | # aas | # FBs | # codons |
|----------|-------|-------|----------|
| Quartets | 5     | 5     | $\varphi(5)=4$ |
| Doublets | 9     | 6     | $\varphi(6)=\varphi(\varphi(9))=2$ |
| Singlets | 2     | 2     | $\varphi(2)=1$ |
| Triplet  | 1     |       | $\varphi(6)+\varphi(2)=2+1=3$ |
| sextets  | 3     |       | $\varphi(6)+\varphi(5)=2+4=6$ |

We have therefore found the correct multiplet structure and the correct degeneracy to get the 61 coding codons:

$$(5\times4+9\times2+2\times1)+(1\times3+3\times6)=61 \qquad (8)$$

**3. Some consequences**

Adding the three stops to equation (8) gives the 64 required codons. Strikingly, this latter number is also equal to the number of numbers relatively prime to our starting number 192 because we have φ(192)=64. In the last Section, we have seen that the degeneracy class-numbers are given by the φ-function which gives the number of coprimes of a given number. Now, we consider also these coprimes, themselves. We have

$$\varphi(5): \{1, 2, 3, 4\}; \sigma^{(5)}=10 \quad (9)$$

$$\varphi(9): \{1, 2, 4, 5, 7, 8\}; \sigma^{(9)}=27 \quad (10)$$

$$\varphi(6): \{1, 5\}; \sigma^{(6)}=6 \quad (11)$$

$$\varphi(2): \{1\}; \sigma^{(2)}=1 \quad (12)$$

where $\sigma^{(n)}$ is the sum of the members of $\varphi(n)$. Also, for the triplet and the sextets, we could compute the corresponding $\sigma$-sums

$$\sigma^{(2)}+\sigma^{(6)}=1+6=7 \quad (13)$$

$$\sigma^{(5)}+\sigma^{(6)}=10+6=16 \quad (14)$$

From Eqs.(9)-(12) we get 44 which equals the total number of degenerate codons (41) and 3 stops and this could be easily established by separating the only "3" from the other twelve numbers in the four sets: 3+41=44. Now, from Eqs.(13)-(14), 7+16=23. This last number fits perfectly our findings in the preceding Section. First, the quartets, the doublets and the singlets whose degeneracy class-numbers are given by *one* $\varphi$-function contain exactly 16 amino acids. Second, the triplet and *two-times* 3 sextets ($S^{IV}$, $R^{IV}$, $L^{IV}$, $S^{II}$, $R^{II}$, $L^{II}$) gives 7 objects. Note that the degeneracy class-numbers of the last two cases are given by a linear combination of the *two* $\varphi$-functions. The relation 23=16+7 has also been found in the context of our model of the genetic code based on 23! (see the introduction and Négadi, 2009). We could also take the two sets (9)-(14) together and obtain

$$23+44=67 \quad (15)$$

Interestingly, and re-interpreting, we get again (see Négadi, 2009) a number which coincides with the total number of carbon atoms in the 20 amino acids in agreement with Račocević's Cyclic Invariant Periodic System (CIPS) classification (Rakočević's, 2009) where there are 23 carbon atoms in the *Primary SuperClass* (PSC) and 44 carbon atoms in the *Secundary SuperClass* (SSC). Several other carbon atom patterns could be obtained by arranging the terms as, for example, 33+34, 28+39, 30+37, etc. (see Négadi, 2006, 2009) which all describe physicochemical known partitionings, including the recent 3D tetrahedron model of the genetic code by Filatov (Filatov, 2009). Concerning this latter model, we could even deduce the following identities 33+34=32+35=67, at the carbon atom level (see Négadi, 2011). As another possible consequence, let us take the equations (9)-(12), *only*. The even sums give $\sigma^{(5)}+\sigma^{(6)}=10+6=16$ and the odd sums $\sigma^{(9)}+\sigma^{(2)}=27+1=28$. These latter two numbers are two characteristic numbers in the 28-gon polyhedral model of the genetic code by Yang (Yang, 2003). They could be made equal respectively to the number of vertices and the number of faces verifying Euler's formula 16+28-42=2, with 42 being the number of edges. As a closing remark, let us return to our starting point, the number 192. We have already explained that this number is equal to the total number of nucleotides in the 64 *RNA* (triplets) codons. Seemingly, nothing would prevent us from taking 64 *DNA* codon, instead, and, why not, both. This latter case where the staring number is 384=192+192 is very interesting as it nicely generalizes the one based on 192. There is only one supplementary factor, 2, and adding its prime index, 1, with sum **2+1** and we could deduce easily the new multiplet structure, (P)' (see the first Section):

$$2+2+2+1+1+1=9 \text{ "doublets"}$$

$$2+2+1=5 \text{ "quartets"}$$

$$1 \text{ "triplet"} \qquad\qquad (P)'$$

$$2 \text{ "singlets"}$$

$$6=3+(\mathbf{2+1}) \text{ "sextets" } (S^{IV}, R^{IV}, L^{IV})+(\{\mathbf{S^{II}, R^{II}}\}, \mathbf{L^{II}})$$

$$2+1=3 \text{ "stops"}$$

Comparison with (P), shows that the supplementary term (in bold) completes the picture with a total of 26 objects: 23 Amino Acids Signals (17 amino acids with no degeneracy at the first-base position and the three amino acids doubly-degenarate at the first-codon position or 6 objects: ($S^{IV}$, $R^{IV}$, $L^{IV}$)+( $S^{II}$, $R^{II}$; $L^{II}$), on the one hand, and 3 stops, on the other. Interestingly, a link between the method, used in this paper, and the one relying on the number 23! (Négadi, 2007, 2009) could be seen as follows. First, from the results of Section 2 and the above pattern (P)' (discarding the 3 "stops"), we have that the number of amino acids computed using the φ-function, Eqs.(3)-(5), is equal to 16 (9+5+2) and the remaining 7 are those for the triplet and the six "sextets" (obtained as linear combinations) with a total of 23. It appears that the number 23!, equal to 25852016738884976644000, has 16 even digits and 7 odd digits. Second, another link between the numbers 23, 23! and the number 192, which is at the basis of the present work, could be seen as follows. The number 23! has 192000 divisors, or $\tau(23!)=192000$ where the τ-function counts the number of divisors. Now, we know that the number 1255 which, physically, is the number of nucleons in the 20 amino acids side-chains and, mathematically, it corresponds to the number of partitions of the number 23, *i.e.*, numbpart(23)=1255 which is also the number of irreducible representations of the symmetric group $S_{23}$ (see Négadi, 2007). We have $\tau(23!)/\varphi(\text{numbpart}(23))=192$, where we used Euler's φ-function. The number 23, which is considered by many people more important than the number 20, could be viewed several ways: 1) 20 amino acids and 3 stops, 2) 17 amino acids with no "degeneracy" at the first codon-position and 6 "sextets (ser$^{II, IV}$; leu$^{II, IV}$; arg$^{II, IV}$), degenerated at the first codon-position and 3) 22 amino acids (including Selenocysteine and Pyrrolysine) and 1 stop signal[1].

The numbers 192 and 384 seem very important numbers. For example they are basic numbers in the Pythagorean music tuning system. For visualization, we reproduce below an extract of the Pythagorean system series (frequencies in Hz)

| | | | | | | | | |
|---|---|---|---|---|---|---|---|---|
| F | | | | 5+ | 11- | 21+ | 43- | 85+ |
| C | 2 | 4 | 8 | 16 | 32 | 64 | 128 | 256 |
| G | 6 | 12 | 24 | 48 | 96 | **192** | **384** | 768 |
| D | 18 | 36 | 72 | 144 | 288 | 576 | | |
| A | 54 | 18 | 216 | 432 | 864 | | | |
| E | 162 | 324 | 648 | | | | | |
| B | 486 | | | | | | | |

---

[1] I thank Branko Dragovich for his comments on the number 23.

It is said that Plato (the soul of the World) choosed 192 as a starting point while Timaeus the Locrian choosed 384. Anyway we have, first, an horizontal *octave-doubling* in the rows which has been paralleled with the process of cell division in nature. Second, in all columns (excluding here the first row, F), the terms are multiplied by 3 when going down but, interestingly, we have also a φ-function iterative *action*: the φ-function of each term gives the preceding one. For example, in the 6[th] column φ(576)=192; φ(192)=φ(φ(576))=64. We remind that the φ-function has played a great role in this work. Finally, 384 is also the total number of atoms in the 20 amino acids (side-chain and bloc) as well as the total number of nucleotides in 64 RNA-codons and in 64 DNA codons (see above).

**Acknowledgement**

I warmly acknowledge Professor Branko Dragovich for his interesting remarks and comments on this work. Some improvements of the manuscript where born while exchanging our ideas.